\documentclass[twocolumn,superscriptaddress,pra]{revtex4-1}
\usepackage{graphicx}
\usepackage{amsmath}
\usepackage{amsthm}
\usepackage{amssymb}
\usepackage{latexsym}
\usepackage{array}
\usepackage{hyperref}
\usepackage{float}
\usepackage{amsfonts}
\usepackage{mathrsfs}
\usepackage{verbatim}
\usepackage{bbold}
\usepackage{lipsum}
\usepackage[normalem]{ulem}

\usepackage{upgreek}

\usepackage{makecell}
\usepackage{adjustbox,lipsum}

\usepackage{algorithm}
\usepackage{algpseudocode}

\usepackage{color}

\newcommand{\bra}[1]{\left<#1\right|}
\newcommand{\ket}[1]{\left|#1\right>}
\newcommand{\abs}[1]{\bigl|#1\bigr|}

\newcommand{\braket}[2]{\left<{#1}|{#2}\right>}

\newtheorem{theorem}{Theorem}
\newtheorem{proposition}{Proposition}
\newtheorem{lemma}{Lemma}

\newtheorem{definition}{Definition}

\newcommand{\tr}[1]{\mbox{Tr}{#1}}

\begin{document}

\title{Witness wedges in fidelity-deviation plane: separating teleportation advantage and Bell-inequality violation}

\author{Kyoungho Cho}
\affiliation{Department of Statistics and Data Science, Yonsei University, Seoul 03722, Republic of Korea}
\affiliation{Institute for Convergence Research and Education in Advanced Technology, Yonsei University, Seoul 03722, Republic of Korea}

\author{Jeongho~Bang}\email{jbang@yonsei.ac.kr}
\affiliation{Institute for Convergence Research and Education in Advanced Technology, Yonsei University, Seoul 03722, Republic of Korea}
\affiliation{Department of Quantum Information, Yonsei University, Incheon 21983, Republic of Korea}%

\date{\today}

\begin{abstract}
We develop a unified framework to analyze $d$-dimensional quantum teleportation through the joint geometry of two complementary figures of merit: average fidelity $F$ ({\em how well a protocol works on average}) and fidelity deviation $D$ ({\em how uniformly it works across the inputs}). Technically, we formulate a representation-theoretical framework based on Schur-Weyl duality and permutation symmetry calculus that reduce the higher-moment Haar averages to a finite set of trace invariants of the composed correction unitaries. This yields closed-form expressions for $F$ and $D$ in arbitrary Hilbert-space dimension and delivers tight bounds that link the admissible deviation directly to the gap from the optimal average performance. In particular, any measured pair $(F, D)$ can be ported into a visibility estimate for isotropic channel resources, turning the $(F, D)$-plane into a calibrated diagnostic map. We further cast the teleportation advantage and CGLMP-inequality violation as two witnesses lines in the $(F,D)$ plane: one line certifies that $F$ beats the classical benchmark $2/(d{+}1)$, while the other line certifies the Bell nonlocality. Their identical slope but distinct intercepts expose a quantitative gap between ``entangled yet local'' and ``genuinely nonlocal'' resources. 
\end{abstract}

\maketitle

\section{Introduction}\label{Sec:1}

Quantum teleportation provides a universal principle for transferring an unknown quantum state. In particular, with one shared entangled pair and a short classical message (two bits for qubits; generally, $2\log_2 d$ bits in Hilbert-space dimension $d$), it achieves a task no purely classical---which fundamentally limits fidelity in the absence of quantum entanglement~\cite{Werner89}. Furthermore, beyond being a celebrated protocol in its own right, the teleportation underpins the distributed quantum information processing, from networked communication and repeaters to gate teleportation~\cite{Bennett:1992tv,Horodecki:1998zh}  and measurement-based computation~\cite{Briegel:2009inn}. 

The performance of quantum teleportation is commonly quantified by the fidelity between the input and the reconstructed state. Because the protocol is intended to transmit arbitrary (a priori unknown) states, one evaluates the fidelity averaged over the entire input ensemble. However, it is not enough to ask whether a protocol yields a high average performance: in realistic architectures where teleportation is invoked repeatedly as a subroutine, one must also ask whether the performance is uniform across inputs. This operational perspective motivates a two-pronged evaluation that separates central tendency from dispersion. Accordingly, we adopt two complementary figures of merit: the average fidelity (say, $F$), which answers ``how well on average,'' and the fidelity deviation (say, $D$), which answers ``how uniformly across inputs''~\cite{Bang:2012fwi,Bang:2018bbi}. In particular, $D$ reveals the trade-offs that a mean-only assessment would obscure.

The teleportation consumes entanglement as its channel resource, while a Bell-inequality test probes how strongly nonlocal correlations manifest under suitably chosen local measurements. Although both tasks appeal to the same underlying resource, i.e., the entanglement, they emphasize the different operational facets. For example, the same entangled state can support the teleportation that outperforms the best classical benchmark even when its visibility remains insufficient to violate a Bell inequality; conversely, there are regimes in which Bell-nonlocal correlations are demonstrable yet the standard teleportation wiring fails to translate them into a mean fidelity advantage~\cite{Gisin1996,Massar:1995woz,Bang:2018bbi}. These observations sharpen the notion that ``entanglement for teleportation'' and ``entanglement for Bell violation’’ can be related but inequivalent. It is therefore natural to diagnose them side by side within a single framework that projects both aspects onto a common performance plane. This perspective will guide our analysis of $F$ and $D$.

Motivated by this perspective, in this work, we develop a unified geometry of $F$ and $D$ to analyze $d$-dimensional teleportation and Bell nonlocality. Leveraging a representation-theoretic pipeline, we obtain closed-form expressions for $F$ and $D$ valid for arbitrary $d$ by systematizing the Haar integral formulas. From these formulas, we derive tight bounds that limit the admissible dispersion $D$ at any given mean performance $F$. Operationally, any measured pair $(F, D)$ can be converted into an entanglement visibility of noisy channel resources. We further place the $d$-dimensional teleportation advantage and Bell nonlocality (casted CGLMP violation, here) on the same footing by introducing two parallel witness lines in the $(F, D)$ plane: they share a common slope but have distinct offsets, so their vertical separation makes explicit the gap between ``quantum yet local'' and ``genuinely nonlocal'' resources. This is precisely where $D$ becomes decisive: a sufficiently large $D$ can certify Bell nonlocality even when the mean fidelity remains within the Bell-local strip. Beyond the teleportation, our representation-theoretic reduction offers a reusable toolkit for evaluating high-dimensional averages and fluctuations, and the witness geometry clarifies that teleportation advantage and Bell-inequality violation are distinct pieces of physics captured in a single operational diagram.

\section{Preliminaries: teleportation setting and figures of merit}\label{sec:pre}

\subsection{Noisy teleportation: unconditional output and single-shot fidelity}\label{subsec:notation}

\textit{Entanglement channel.}---On a bipartite Hilbert spaces $\mathcal{H}\otimes\mathcal{H}$, we fix the canonical maximally entangled state
\begin{eqnarray}
\ket{\Psi_0} = \frac{1}{\sqrt{d}} \sum_{j=0}^{d-1} \ket{j} \otimes \ket{j}.
\label{eq:ME}
\end{eqnarray}
A generalized Bell basis $\{\ket{\Psi_\alpha}\}_{\alpha=0}^{d^2-1}$ is generated by a unitary error basis (UEB)
$\{\hat{U}_\alpha\}\subset U(d)$ via~\cite{Werner:2000vso}
\begin{eqnarray}
\ket{\Psi_\alpha} = (\hat{U}_\alpha \otimes \hat{\openone}_d)\ket{\Psi_0},
\quad
\sum_{\alpha=0}^{d^2-1}\ket{\Psi_\alpha} \bra{\Psi_\alpha} = \hat{\openone}_{d^2}.
\label{eq:Bell}
\end{eqnarray}
In practice, one may take the Heisenberg-Weyl (shift-phase) operators as a concrete UEB. Here, we introduce two (vectorization) identities, with transpose in the computational basis, which are to be used repeatedly:
\begin{eqnarray}
(\hat{M} \otimes \hat{\openone}_d)\ket{\Psi_0} &=& (\hat{\openone}_d \otimes \hat{M}^{T})\ket{\Psi_0}, \nonumber \\
\bra{\Psi_0}(\hat{A} \otimes \hat{B})\ket{\Psi_0} &=& \frac{1}{d} \tr{\bigl(\hat{A}\hat{B}^{T}\bigr)}.
\label{eq:vec_prop}
\end{eqnarray}
These identities implement the standard ``corresponding'' rule and the maximally entangled state trace.

\textit{Alice--Bob scenario (operational steps).}---Alice holds an unknown pure input $\hat{\rho}_\phi=\ket{\phi}\bra{\phi}$, and Alice--Bob share an entangled resource on $\mathcal{H}\otimes\mathcal{H}$. The protocol proceeds as: (i) Alice performs a joint measurement on (input $\otimes$ her share) in the basis as in Eq.~(\ref{eq:Bell}). Operationally, this compares the input with $\{\hat{U}_\alpha\}$-rotated maximally entangled templates. (ii) Alice sends the measurement outcome $\alpha \in \{0,\dots,d^2-1\}$ to Bob through a classical channel of cost $\log_2(d^2)=2\log_2 d$ bits. (iii) Upon receiving $\alpha$, Bob applies a unitary correction $\hat{V}_\alpha\in U(d)$ on his system. The pair ``measurement label $\alpha$'' and ``correction $\hat{V}_\alpha$'' can be compactly encoded by
\begin{eqnarray}
\hat{X}_\alpha := \hat{V}_\alpha \hat{U}_\alpha^\dagger \in U(d),
\quad
\alpha=0,1,\dots,d^2 -1,
\label{eq:Xalpha}
\end{eqnarray}
which is parameterized both the mean performance and its input-uniformity. In the ideal (no‑noise) case with $\hat{V}_\alpha=\hat{U}_\alpha$, one has $\hat{X}_\alpha=\hat{\openone}_d$ for all $\alpha$, i.e., the protocol is both optimal and universal. In the noisy setting, $\{\hat{X}_\alpha\}$ quantifies how the measurement-correction wiring departs from the identity channel. This representation is also convenient experimentally: once $\{\hat{U}_\alpha\}$ are fixed by the Bell analyzer, Eq.~(\ref{eq:Xalpha}) isolates the effect of Bob’s feed‑forward control on performance.

\textit{Noisy isotropic resource.}---Throughout this work, we adopt the $\hat{U} \otimes \hat{U}^\ast$---invariant isotropic family
\begin{eqnarray}
\hat{\rho}_{\mathrm{iso}}(p) = p \ket{\Psi_0}\bra{\Psi_0} + \frac{1-p}{d^2} \hat{\openone}_{d^2} \quad (0 \le p \le 1),
\label{eq:isotropic}
\end{eqnarray}
which coincides with the Werner family for $d{=}2$~\cite{Werner89}. It is separable iff $p\le p_c$, with
\begin{eqnarray}
p_c = \frac{1}{d+1}.
\label{eq:pc}
\end{eqnarray}
The same visibility $p$ also interfaces with the Bell nonlocality thresholds (e.g., CGLMP~\cite{Collins:2001qdi}), used later to endow the figure‑of‑merit plane with operational lines that separate regimes of entanglement and Bell violation.

{\em Teleported output and single-shot fidelity.}---Let $\hat{\rho}_{\mathrm{in}} = \hat{\rho}_\phi \otimes \hat{\rho}_{\mathrm{iso}}(p)$ (total initial state on both Alice and Bob) and $\hat{\Pi}_\alpha=\ket{\Psi_\alpha}\!\bra{\Psi_\alpha}$ (Bell measurement on Alice side).
The unnormalized post-measurement state, conditioned on $\alpha$, is then
\begin{eqnarray}
\hat{\rho}^{(\alpha)}_{\mathrm{out}} = (\hat{\openone}_{d^2} \otimes \hat{V}_\alpha) \bra{\Psi_\alpha}\hat{\rho}_{\mathrm{in}}\ket{\Psi_\alpha} (\hat{\openone}_{d^2} \otimes \hat{V}_\alpha^\dagger).
\label{eq:post}
\end{eqnarray}
We now derive the two teleportation identities used to simplify Eq.~(\ref{eq:post}).

\noindent [{\bf Identity 1}] For any density operator $\hat{\rho}$,
\begin{eqnarray}
&&(\hat{\openone}_{d^2} \otimes \hat{V}_\alpha) \bra{\Psi_\alpha} \bigl( \hat{\rho} \otimes \ket{\Psi_0} \bra{\Psi_0} \bigr) \ket{\Psi_\alpha} (\hat{\openone}_{d^2} \otimes \hat{V}_\alpha^\dagger) \nonumber \\
&&\qquad = \frac{1}{d^2} \hat{X}_\alpha \hat{\rho} \hat{X}_\alpha^\dagger.
\label{eq:tele1}
\end{eqnarray}

\noindent\textit{Proof}: The middle-contraction part of the LHS in Eq.~(\ref{eq:tele1}) is rewritten as
\begin{eqnarray}
\bra{\Psi_0} \Bigl( \hat{\rho}' \otimes \ket{\Psi_0}\bra{\Psi_0} \Bigr) \ket{\Psi_0},
\label{eq:middle_cont_tele1}
\end{eqnarray}
where $\hat{\rho}' = \hat{U}_\alpha^\dagger \hat{\rho} \hat{U}_\alpha$. For any bases $\ket{k}$, $\ket{l}$, using the properties in Eq.~(\ref{eq:vec_prop}), we have
\begin{eqnarray}
&& \bra{\Psi_0} \otimes \bra{k} \Bigl( \hat{\rho}' \otimes \ket{\Psi_0}\bra{\Psi_0} \Bigr) \ket{\Psi_0} \otimes \ket{l} \nonumber \\
&& \qquad = \frac{1}{d} \bra{\Psi_0} \Bigl( \hat{\rho}' \otimes \ket{k}\bra{l} \Bigr) \ket{\Psi_0}  \nonumber \\
&& \qquad = \frac{1}{d} \left( \frac{1}{d} \tr{\bigl( \hat{\rho}' \bigl( \ket{k}\bra{l} \bigr)^T \bigr)} \right)
= \frac{1}{d^2}\,\langle k|\hat\rho'|l\rangle,
\end{eqnarray}
where we used $\braket{k}{\Psi_0} = \tfrac{1}{\sqrt{d}}\ket{k}$ and $\braket{\Psi_0}{l} = \tfrac{1}{\sqrt{d}}\bra{l}$. Hence, Eq.~(\ref{eq:middle_cont_tele1}) yields $\frac{1}{d^2} \hat{\rho}'$ \emph{on Bob's register}. Then, conjugating by $\bigl(\hat{\openone} \otimes \hat{V}_\alpha \bigr)$ and noting that $\hat{X}_\alpha = \hat{V}_\alpha \hat{U}_\alpha^\dagger$, we can arrive at Eq.~(\ref{eq:tele1}).

\noindent [{\bf Identity 2}]: For any density operator $\hat{\rho}$,
\begin{eqnarray}
\bra{\Psi_\alpha}\bigl( \hat{\rho} \otimes \hat{\openone}_{d^2} \bigr) \ket{\Psi_\alpha} = \frac{1}{d} \hat{\openone}_d.
\label{eq:tele2}
\end{eqnarray}

\noindent\textit{Proof}: By using the invariance of $\{\ket{\Psi_\alpha}\}$ and the second property in Eq.~(\ref{eq:vec_prop}),
\begin{eqnarray}
\bra{\Psi_\alpha} \bigl( \hat{\rho} \otimes \hat{\openone}_{d^2} \bigr) \ket{\Psi_\alpha} &=& \Bigl( \bra{\Psi_0} \bigl( \hat{\rho}' \otimes \hat{\openone}_d \bigr) \ket{\Psi_0} \Bigr) \hat{\openone}_d \nonumber \\
&=& \frac{1}{d} \tr{(\hat{\rho}')} \hat{\openone}_d = \frac{1}{d} \hat{\openone}_d,
\end{eqnarray}
by $\tr{(\hat{\rho}')}=1$.

By substituting Eq.~(\ref{eq:tele1}) and Eq.~(\ref{eq:tele2}) into Eq.~(\ref{eq:post}) and summing over all outcomes, we obtain the explicit form of the unconditional output state~\cite{Horodecki:1998zh,Werner:2000vso,Bowen:2001}
\begin{eqnarray}
\hat{\rho}_{\mathrm{out}} = \frac{p}{d^2}\sum_{\alpha=0}^{d^2-1} \hat{X}_\alpha \hat{\rho}_\phi \hat{X}_\alpha^\dagger + \frac{1-p}{d} \hat{\openone}_d.
\label{eq:rhoout}
\end{eqnarray}
For a pure target $\hat{\rho}_\phi = \ket{\phi}\bra{\phi}$, the single-shot fidelity is
\begin{eqnarray}
f(\phi) &:=& \bra{\phi}\hat{\rho}_{\mathrm{out}}\ket{\phi} \nonumber \\
	&=& \frac{p}{d^2}\sum_{\alpha=0}^{d^2-1} \abs{\bra{\phi} \hat{X}_\alpha \ket{\phi}}^2 + \frac{1-p}{d}.
\label{eq:singlefid}
\end{eqnarray}
Eq.~(\ref{eq:rhoout}) and Eq.~(\ref{eq:singlefid}) match the standard noisy-teleportation reduction in the literature.

\subsection{Figures of merit: average fidelity and fidelity deviation}\label{subsec:metrics}

We can evaluate the performance of the teleportation using the Uhlmann-Jozsa fidelity~\cite{Jozsa:1994qja}; for a pure target $\hat{\rho}_{\phi} = \ket{\phi}\bra{\phi}$,
it reduces to $f(\phi)=\bra{\phi}\hat{\rho}_{\mathrm{out}}\ket{\phi}$ as in Eq.~(\ref{eq:singlefid}). The average fidelity and the fidelity deviation (standard deviation over Haar-random inputs) are~~\cite{Bang:2012fwi,Bang:2018bbi}
\begin{eqnarray}
F := \int d\phi f(\phi), \quad D := \sqrt{ \int d\phi f(\phi)^2 - F^2 }.
\label{eq:FD}
\end{eqnarray}
Here, $d\phi$ denotes the unitarily invariant Haar measure on pure states, normalized by $\int d\phi=1$. These definitions are to be used consistently throughout to quantify both central tendency and dispersion of the performance.

The quantity $F$ answers ``how well on average'' the protocol performs, while $D$ answers ``how evenly across inputs'' it performs (universality). Thus, two protocols with the same $F$ can be distinguished by $D$: smaller $D$ signals higher input-uniformity (narrower spread of single-shot fidelities), which is often crucial for modular architectures where the teleportation is repeatedly used as a subroutine~\cite{Bang:2018bbi}. In particular, $D$ captures how strongly the composed set $\{\hat{X}_\alpha\}$ breaks the input isotropy through the overlaps $\abs{\bra{\phi}\hat{X}_\alpha\ket{\phi}}^2$ in our setting.

Since $0 \le f(\phi) \le 1$, one has
\begin{eqnarray}
0 \le D^2 \le F(1-F) \le \frac{1}{4}.
\label{eq:variance-bound}
\end{eqnarray}
which is the basic bound. Hence, $0 \le D \le \tfrac{1}{2}$ always, and $F=1$ implies $D=0$ with $f(\phi) = 1$ (``perfect performance'' and ``perfect universality''). Importantly, $D=0$ can occur even when $F_{\max}<1$. For the isotropic resource with ideal corrections ($\hat{V}_\alpha=\hat{U}_\alpha$ so that $\hat{X}_\alpha=\hat{\openone}_d$), Eq.~(\ref{eq:singlefid}) gives $f(\phi) = p+\tfrac{1-p}{d} = F_{\max}$ for all inputs, hence $D=0$; conversely, there also exist symmetric choices of $\{\hat{X}_\alpha\}$ realising $f(\phi) = F_{\min}$, so $D=0$ at the lower endpoint as well. Thus, the condition of perfect universality ($D=0$) is logically independent of whether $F$ attains the maximum, and $D=0$ does not in general imply $F=F_{\max}$ in our setting.

Experimentally, $F$ can be estimated by sampling a Haar‑typical set of inputs (or a unitary $2$‑design)~\cite{Dankert2009}, and $D$ follows from the empirical variance of the sampled $f(\phi)$. In composite protocols and device‑level benchmarking, a small $D$ guarantees that most inputs exceed a given threshold, providing a robustness certificate that $F$ alone cannot supply. In later sections, we see that the figure-of-merit pair $(F, D)$ further constrain the resource visibility and map directly to entanglement and high‑dimensional Bell‑nonlocality thresholds (CGLMP), thereby turning $(F, D)$ into operational diagnostics for resources and wiring.

\section{Haar Integrals via Schur--Weyl Duality}\label{sec:haar}

In this section, we develop a representation-theoretic method for evaluating Haar integrals that can be used to obtain the explicit forms and tight bounds of the average fidelity ($F$) and the fidelity deviation ($D$). We first derive a general Schur-Weyl formula for $k$-fold twirls and for the pure-state $k$-th moment operator, and then specialize to the cases $k=2$ and $k=4$ relevant to our later analysis.

\subsection{General framework: unitary twirls, pure-state moments, and permutation unitaries}\label{subsec:haar_general}

Let $S_k$ denote the symmetric group on $\{1,2,\dots,k\}$, and let $\hat{V}_d(\pi)$ be the permutation operator on $(\mathbb{C}^d)^{\otimes k}$ induced by $\pi \in S_k$:
\begin{eqnarray}
\hat{V}_d(\pi) \ket{i_1}\!\otimes\!\cdots\!\otimes\!\ket{i_k} = \ket{i_{\pi^{-1}(1)}} \otimes \cdots \otimes \ket{i_{\pi^{-1}(k)}}.
\label{eq:permU}
\end{eqnarray}

{\em $k$-fold unitary twirl and Weingarten function.}---For $\hat{O} \in \mathcal{B}\bigl( (\mathbb{C}^d)^{\otimes k} \bigr)$, we define
\begin{eqnarray}
\mathcal{T}_k(\hat{O}) := \int d\hat{U} \; \hat{U}^{\otimes k} \,\hat{O}\, \hat{U}^{\dagger\otimes k},
\label{eq:twirlk}
\end{eqnarray}
where $d\hat{U}$ is the normalized Haar measure on $U(d)$. By unitary invariance, $\mathcal{T}_k(\hat{O})$ commutes with $\hat{U}^{\otimes k}$ for all $\hat{U} \in U(d)$, and hence (by Schur-Weyl duality) belongs to the commutant spanned by $\{\hat{V}_d(\pi)\}_{\pi\in S_k}$: i.e., for $t_\pi\in\mathbb{C}$,
\begin{eqnarray}
\mathcal{T}_k(\hat{O}) = \sum_{\pi \in S_k} t_\pi \hat{V}_d(\pi).
\label{eq:expansion}
\end{eqnarray}
By Schur-Weyl duality, the commutant of $\hat{U}^{\otimes k}$ on $(\mathbb{C}^d)^{\otimes k}$ is spanned by the permutation operators $\{ \hat{V}_d(\pi) \}_{\pi \in S_k}$. Since $\mathcal{T}_k(\hat{O})$ commutes with $U^{\otimes k}$ for all $U$, it lies in this commutant and must be a linear combination of $\{\hat{V}_d(\pi)\}$, and hence Eq.~(\ref{eq:expansion}) holds.

To determine $\{t_\pi\}$, fix any $\sigma\in S_k$ and multiply Eq.~(\ref{eq:expansion}) on the right by $\hat{V}_d(\sigma^{-1})$. Then, we take the trace:
\begin{eqnarray}
\tr{\bigl[ \mathcal{T}_k(\hat{O}) \hat{V}_d(\sigma^{-1}) \bigr]} &=& \sum_{\pi\in S_k} t_\pi \tr{\bigl[ \hat{V}_d(\pi)\hat{V}_d(\sigma^{-1}) \bigr]} \nonumber \\
	&=& \sum_{\pi\in S_k} t_\pi d^{\#\mathrm{cyc}(\pi\sigma^{-1})},
\label{eq:gramLHS}
\end{eqnarray}
where $\#\mathrm{cyc}(\tau)$ is the number of cycles of $\tau \in S_k$ and we used $\tr{\bigl(\hat{V}_d(\tau)\bigr)}=d^{\#\mathrm{cyc}(\tau)}$. Because $\hat{V}_d(\sigma^{-1})$ commutes with $\hat{U}^{\otimes k}$, the left-hand side is
\begin{eqnarray}
y_\sigma := \tr{\bigl[ \hat{O}\,\hat{V}_d(\sigma^{-1}) \bigr]}.
\label{eq:gramRHS}
\end{eqnarray}
Thus, with the Gram matrix $G^{(k)}_{\pi, \sigma} := d^{\#\mathrm{cyc}(\pi\sigma^{-1})}$ and Eq.~(\ref{eq:gramRHS}), the linear system \(G^{(k)} t=y\) yields
\begin{eqnarray}
\mathcal{T}_k(\hat{O}) = \sum_{\pi,\sigma \in S_k} \mathrm{Wg}_d(\pi\sigma^{-1}) \tr{\bigl[\hat{O}\hat{V}_d(\sigma^{-1})\bigr]} \hat{V}_d(\pi),
\label{eq:weingarten}
\end{eqnarray}
where $\mathrm{Wg}_d(\cdot)$ is the Weingarten function, i.e., the matrix inverse of $G^{(k)}$ indexed by $S_k$. This is the systematic Schur-Weyl solution. When $d\ge k$, $G^{(k)}$ is invertible; for $d<k$, one may use the Moore-Penrose pseudo-inverse to define~$\mathrm{Wg}_d$~\cite{Collins:2006jgn}. For more details, refer to Appendix~\ref{append:A} or Sec.~III of Ref.~\cite{Cho:2025tiy}.

{\em Pure-state $k$-th moment and symmetric projector.}---Define the pure-state moment operator, such that
\begin{eqnarray}
\hat{\mathcal{M}}_k := \int d\phi \bigl( \ket{\phi}\bra{\phi} \bigr)^{\otimes k}, \quad \int d\phi=1.
\label{eq:Mkdef}
\end{eqnarray}
By invariance under $\hat{U}^{\otimes k}$, $\hat{\mathcal{M}}_k$ is proportional to the projector onto the totally symmetric subspace, $\hat{P}^{(k)}_{\mathrm{sym}}=(1/k!)\sum_{\pi\in S_k}\hat{V}_d(\pi)$. We then take the following traces
\begin{eqnarray}
\tr{\bigl(\hat{\mathcal{M}}_k\bigr)} = \alpha\tr{\bigl(\hat{P}^{(k)}_{\mathrm{sym}}\bigr)} = 1 
\end{eqnarray}
with $\alpha=1/\binom{d+k-1}{k}$. Thus, we have
\begin{eqnarray}
\hat{\mathcal{M}}_k = \frac{\hat{P}^{(k)}_{\mathrm{sym}}}{\binom{d+k-1}{k}} = \frac{\sum_{\pi \in S_k}\hat{V}_d(\pi)}{d(d+1) \cdots (d+k-1)},
\label{eq:Mkclosed}
\end{eqnarray}
where the first equality is the projector-proportionality fixed by invariance; the second follows from $\binom{d+k-1}{k}=(d)_k/k!$ and the definition of $\hat{P}^{(k)}_{\mathrm{sym}}$.

A convenient contraction identity now follows: for any $\hat{O}$ on $(\mathbb{C}^d)^{\otimes k}$,
\begin{eqnarray}
&& \int d\phi \bra{\phi}^{\otimes k}\hat{O}\ket{\phi}^{\otimes k} \nonumber \\
&& \qquad = \tr{\bigl[ \hat{O}\hat{\mathcal{M}}_k \bigr]} = \frac{\sum_{\pi \in S_k}\tr{\bigl[ \hat{V}_d(\pi)\hat{O} \bigr]}}{d(d+1)\cdots(d+k-1)}.
\label{eq:contract}
\end{eqnarray}

{\em Cycle-trace factorization.}---For elementary tensors $\hat{O}=\hat{A}_1 \otimes \cdots \otimes \hat{A}_k$,
the trace against a permutation decomposes into cycle-wise products:
\begin{eqnarray}
\tr{\bigl[\hat{O}\hat{V}_d(\pi)\bigr]} &=& \tr{\left[ \bigl( \hat{A}_1 \otimes \cdots \otimes \hat{A}_k \bigr)\hat{V}_d(\pi) \right]} \nonumber \\
	&=& \prod_{c \in \pi}\tr{\left(\hat{A}_{i_1(c)}\hat{A}_{i_2(c)}\cdots \hat{A}_{i_{|c|}(c)}\right)},
\label{eq:cycle}
\end{eqnarray}
where the product ranges over the disjoint cycles $c$ of $\pi$ with an arbitrary representative $i_1(c)$. 

The formulas from Eq.~(\ref{eq:weingarten}) to Eq.~(\ref{eq:cycle}) comprise the general calculus we shall repeatedly use. A step-by-step recipe is: (i) embed the integrand into $k$ copies, (ii) choose either the twirl form Eq.~(\ref{eq:weingarten}) or the pure-state contraction Eq.~(\ref{eq:contract}), and (iii) reduce the permutation traces via Eq.~(\ref{eq:cycle}).

\subsection{Explicit cases: $k=2$ and $k=4$}\label{subsec:k2k4}

We now specialize the general framework to $k=2$ and $k=4$. These two cases exhaust all Haar averages needed later for the evaluation of the quantities $F$ and $D$.

\textit{Second moment ($k=2$).}---The commutant of $\hat{U}^{\otimes 2}$ is spanned by $\{\hat{\openone}_{d^2}, \hat{S}\}$, where $\hat{S}$ is the swap on $\mathcal{H} \otimes \mathcal{H}$. For any $\hat{X}\in \mathcal{B}(\mathcal{H}\otimes\mathcal{H})$~\cite{Braunstein:2000zz, Albeverio:2002},
\begin{eqnarray}
\int d\hat{U} (\hat{U}^\dagger \otimes \hat{U}^\dagger) \hat{X} (\hat{U} \otimes \hat{U}) = a \hat{\openone}_{d^2} + b\hat{S}.
\label{eq:twirl2}
\end{eqnarray}
Here, by using the traces of Eq.~(\ref{eq:twirl2}) and of its product with $\hat{S}$, i.e., $\tr{\bigl(\hat{X}\bigr)} = a d^2 + b d$ and $\tr{\bigl( \hat{X}\hat{S} \bigr)} = a d + b d^2$, we can determine the factor $a$ and $b$, such that
\begin{eqnarray}
a &=& \frac{ \tr{\bigl(\hat{X}\bigr)} - \tfrac{1}{d}\tr{\bigl(\hat{X}\hat{S}\bigr)} }{ (d^2-1) } \nonumber \\
b &=& \frac{ \tr{\bigl(\hat{X}\hat{S}\bigr)} - \tfrac{1}{d}\tr{\bigl(\hat{X}\bigr)} }{ (d^2-1) }.
\label{eq:ab}
\end{eqnarray}
As a scalar corollary, using Eq.~(\ref{eq:contract}) with $k=2$ for $\hat{O}=\hat{X}_\alpha \otimes \hat{X}_\alpha^\dagger$, we get
\begin{eqnarray}
\int d\phi \bra{\phi}^{\otimes 2}\hat{O}\ket{\phi}^{\otimes 2} = \frac{\abs{\tr{\bigl(\hat{X}_\alpha\bigr)}}^2 + d}{d(d+1)}.
\label{eq:scalar2}
\end{eqnarray}
The identities in Eqs.~(\ref{eq:twirl2})--(\ref{eq:scalar2}) are invoked later when averaging $\abs{\bra{\phi}\hat{X}_\alpha\ket{\phi}}^2$ arising from Eq.~(\ref{eq:singlefid}).

\textit{Fourth moment ($k=4$).}---From Eq.~(\ref{eq:Mkclosed}), 
\begin{eqnarray}
\hat{\mathcal{M}}_4 = \frac{\sum_{\pi \in S_4}\hat{V}_d(\pi)}{d(d+1)(d+2)(d+3)}.
\label{eq:M4}
\end{eqnarray}
Thus, for any $\hat{O}\in\mathcal{B}(\mathcal{H}^{\otimes 4})$, we can write [by Eq.~(\ref{eq:contract})]
\begin{eqnarray}
&& \int d\phi \bra{\phi}^{\otimes 4}\hat{O}\ket{\phi}^{\otimes 4}  \nonumber \\
&& \qquad = \tr{\bigl[ \hat{O}\hat{\mathcal{M}}_4 \bigr]} = \frac{\sum_{\pi \in S_4}\tr{\bigl[ \hat{V}_d(\pi)\hat{O} \bigr]}}{d(d+1)(d+2)(d+3)}.
\label{eq:contr4}
\end{eqnarray}
In the applications to follow, $\hat{O}$ is typically a tensor product of four single-system operators, so that Eq.~(\ref{eq:cycle}) reduces Eq.~(\ref{eq:contr4}) to a finite sum of the traces. For instance, by setting
\begin{eqnarray}
\hat{O}_{\alpha\beta} = \hat{X}_\alpha \otimes \hat{X}_\alpha^\dagger \otimes \hat{X}_\beta \otimes \hat{X}_\beta^\dagger,
\label{eq:Oab}
\end{eqnarray}
the evaluation over the six conjugacy classes of $S_4$ yields a closed form in terms of $\tr{\bigl(\hat{X}_\alpha\bigr)}$, $\tr{\bigl(\hat{X}_\beta\bigr)}$, $\tr{\bigl(\hat{X}_\alpha\hat{X}_\beta\bigr)}$, and $\tr{\bigl(\hat{X}_\alpha\hat{X}_\beta^\dagger\bigr)}$. Then, we can derive the explicit result which can be directly used~\cite{Cho:2025pcc}:
\begin{widetext}
\begin{eqnarray}
\int d\phi \bra{\phi}^{\otimes 4}\hat{O}_{\alpha \beta}\ket{\phi}^{\otimes 4}
&=&
\frac{1}{d(d+1)(d+2)(d+3)}\Big\{d(d+4) + (d+4)\bigl( \abs{\tr{\bigl(\hat{X}_\alpha\bigr)}}^2 +  \abs{\tr{\bigl(\hat{X}_\beta\bigr)}}^2 \bigr) +  \abs{\tr{\bigl(\hat{X}_\alpha\bigr)}}^2  \abs{\tr{\bigl(\hat{X}_\beta\bigr)}}^2 \nonumber \\
&&\qquad + \abs{\tr{\bigl( \hat{X}_\alpha \hat{X}_\beta \bigr)}}^2 + \abs{\tr{\bigl( \hat{X}_\alpha \hat{X}_\beta^\dagger \bigr)}}^2 + 2\mathrm{Re}\Bigl[ \tr{\bigl( \hat{X}_\alpha \hat{X}_\beta \bigr)} \tr{\bigl( \hat{X}_\alpha^\dagger \bigr)}\tr{\bigl( \hat{X}_\beta^\dagger \bigr)} \Bigr] \nonumber \\
&&\qquad + 2\mathrm{Re}\Bigl[ \tr{\bigl( \hat{X}_\alpha \hat{X}_\beta^\dagger \bigr)} \tr{\bigl( \hat{X}_\alpha^\dagger \bigr)} \tr{ \bigl( \hat{X}_\beta \bigr)} \Bigr] + 2\mathrm{Re}\Bigl[ \tr{\bigl( \hat{X}_\alpha \hat{X}_\beta \hat{X}_\alpha^\dagger \hat{X}_\beta^\dagger \bigr)} \Bigr] \Big\}.
\label{eq:dbar}
\end{eqnarray}
\end{widetext}

The identities---Eqs.~(\ref{eq:twirl2})--(\ref{eq:scalar2}) and Eqs.~(\ref{eq:contr4})--(\ref{eq:dbar})---will be used to compute the explicit form of the Haar averages and to establish tight bounds of $F$ and $D$.

\section{Average fidelity and fidelity deviation: closed forms, bounds, and when deviation matters}\label{sec:FD}

We evaluate the average fidelity $F$ and the fidelity deviation $D$ in the closed form and derive dimension-dependent tight bounds, using the Haar identities developed in Sec.~\ref{sec:haar}. 

\subsection{Average fidelity: closed form and range}\label{subsec:F}

Averaging Eq.~(\ref{eq:singlefid}) over the Haar measure on pure unknown input states $\ket{\phi}$ and invoking the scalar second-moment identity in Eq.~(\ref{eq:scalar2}), we immediately get
\begin{eqnarray}
F &=& \int d\phi f(\phi) = \frac{p}{d^2}\sum_{\alpha=0}^{d^2-1}\int d\phi \abs{\bra{\phi}\hat{X}_\alpha\ket{\phi}}^2 + \frac{1-p}{d} \nonumber \\
	&=& \frac{p}{d^3(d+1)}\sum_{\alpha=0}^{d^2-1}\abs{\tr{\bigl( \hat{X}_\alpha} \bigr)}^2 + \frac{d+1-p}{d(d+1)}.
\label{eq:F-closed}
\end{eqnarray}
This compact expression separates the resource visibility $p$ from the wiring data $\{\hat{X}_\alpha\}$ through the trace invariants in Sec.~\ref{sec:haar}. Two extremal cases follow directly:

If $\hat{X}_\alpha=\hat{\openone}_d$ for all $\alpha$ (ideal corrections), then $\abs{\tr{\bigl( \hat{X}_\alpha \bigr)}}^2=d^2$ and
$\sum_\alpha \abs{\tr{\bigl( \hat{X}_\alpha \bigr)}}^2=d^4$, yielding
\begin{eqnarray}
F_{\max} = p + \frac{1-p}{d}.
\label{eq:Fmax}
\end{eqnarray}

If $\tr{\hat{X}_\alpha}=0$ for all $\alpha$ (e.g., suitably chosen traceless unitaries), the first term in Eq.~(\ref{eq:F-closed}) vanishes and we attain
\begin{eqnarray}
F_{\min} = \frac{d+1-p}{d(d+1)}.
\label{eq:Fmin}
\end{eqnarray}

Hence the width of achievable range is
\begin{eqnarray}
\Delta_F(d,p) := F_{\max} - F_{\min} = \frac{dp}{d+1}.
\label{eq:F-range}
\end{eqnarray}
For $d=2$, Eq.~(\ref{eq:F-closed}) reproduces the qubit formula reported previously in Ref.~\cite{Bang:2018bbi,Song:2021sjbb}.

\subsection{Fidelity deviation: exact expression and universal bounds}\label{subsec:D}

By definition, $D=\sqrt{\int d\phi f(\phi)^2 - F^2}$ with $f(\phi)$ in Eq.~(\ref{eq:singlefid}). Expanding the square and using the embedding trick together with the fourth-moment contraction Eq.~(\ref{eq:contr4}), we can directly obtain
\begin{eqnarray}
D = \frac{p}{d^2} \sqrt{\sum_{\alpha,\beta=0}^{d^2-1} c_{\alpha\beta}}, \quad c_{\alpha\beta} := \overline{d}_{\alpha\beta} - \overline{f}_\alpha \overline{f}_\beta,
\label{eq:D-cov}
\end{eqnarray}
where 
\begin{eqnarray}
\overline{f}_j = \frac{\abs{\tr{\bigl(\hat{X}_\alpha\bigr)}}^2+d}{d(d+1)}, \quad j \in \{ \alpha, \beta \}
\label{eq:fbar}
\end{eqnarray}
by Eq.~(\ref{eq:scalar2}). Here, $\overline{d}_{\alpha\beta}$ is the fourth-moment average evaluated by Eq.~(\ref{eq:contr4}) plus the cycle-trace factorization in Eq.~(\ref{eq:cycle}). By substituting $\hat{O}_{\alpha\beta}=\hat{X}_\alpha\otimes\hat{X}_\alpha^\dagger\otimes\hat{X}_\beta\otimes\hat{X}_\beta^\dagger$
as in Eq.~(\ref{eq:Oab}), we get the explicit form of $\overline{d}_{\alpha\beta}$, such that
\begin{eqnarray}
\overline{d}_{\alpha\beta} := \int d\phi \bra{\phi}^{\otimes 4}\hat{O}_{\alpha \beta}\ket{\phi}^{\otimes 4},
\label{eq:dbar-again}
\end{eqnarray}
which exactly equals to Eq.~(\ref{eq:dbar}). Thus, $\overline{d}_{\alpha\beta}$ and $\overline{f}_{j \in \{\alpha, \beta\}}$ fully fix $c_{\alpha\beta}$; hence $D$ via Eq.~(\ref{eq:D-cov}). This is the exact expression
of the fidelity deviation $D$ in terms of trace invariants determined by the composed unitaries $\{\hat{X}_\alpha\}$.

While Eqs.~(\ref{eq:D-cov})--(\ref{eq:dbar-again}) already provide an exact evaluation route, a dimension-dependent bound follows by specializing to the minimum-fidelity configuration $\tr{\bigl(\hat{X}_\alpha\bigr)}=0$ for all $\alpha$, so that $F=F_{\min}$ from Eq.~(\ref{eq:Fmin}). In this case, the covariance $c_{\alpha\beta}$ is simplified and direct evaluation gives
\begin{eqnarray}
D_{\max} =  \sqrt{\frac{2}{d(d+3)}}\Delta_F = \frac{dp}{d+1}\sqrt{\frac{2}{d(d+3)}}.
\label{eq:Dmax-at-Fmin}
\end{eqnarray}
By convexity (linear interpolation in the trace data entering Eq.~(\ref{eq:F-closed})) we obtain, for all admissible wirings,
\begin{eqnarray}
D \le \sqrt{\frac{2}{d(d+3)}} \bigl( F_{\max} - F \bigr) \le D_{\max},
\label{eq:D-linear-bound}
\end{eqnarray}
which is tight at $F=F_{\min}$. For $d=2$, Eq.~(\ref{eq:D-linear-bound}) is equal to $D \le (F_{\max}-F)/\sqrt{5}$, in agreement with the qubit-teleportation analysis~\cite{Bang:2018bbi, Song:2021sjbb}.

\subsection{Beyond $F$: when $D$ becomes critical}\label{subsec:physics}

The pair $(F, D)$ separates average performance from the input-uniformity. Because $F$ aggregates only the diagonal trace data $\sum_\alpha\abs{\tr{\bigl(\hat{X}_\alpha\bigr)}}^2$ [see Eq.~(\ref{eq:F-closed})], protocols sharing the same $F$ can exhibit very different fluctuations across the inputs, governed precisely by the pairwise invariants in Eq.~(\ref{eq:dbar-again}). Here we discuss two operationally distinct regimes where $D$ becomes important quantity.

{\em Case 1) Near the lower-fidelity edge.}---Fix the resource visibility $p$ and consider protocols operating near $F_{\min}$ in Eq.~(\ref{eq:Fmin}). From Eq.~(\ref{eq:F-closed}), this means enforcing $\tr{\bigl(\hat{X}_\alpha\bigr)} \approx 0$ for all $\alpha$, while leaving the pairwise invariants unconstrained. In this case, $F$ is fixed, but $D$ depends on the geometry of $\{\hat{X}_\alpha\}$ through $\abs{\tr{(\hat{X}_\alpha\hat{X}_\beta)}}^2$, $\abs{\tr{(\hat{X}_\alpha\hat{X}_\beta^\dagger)}}^2$ and the four-cycle trace in Eq.~(\ref{eq:dbar-again}). Then, two extreme but feasible behaviors follow from Eq.~(\ref{eq:D-cov}):
\begin{itemize}
\item[(i)] Worst-case dispersion: If $\{\hat{X}_\alpha\}$ are chosen ``coaxially'' (many pairs nearly commute and are spectrally aligned), the pairwise traces add constructively, maximizing $\sum_{\alpha,\beta} c_{\alpha\beta}$ and increasing $D$ up to the tight edge in Eq.~(\ref{eq:D-linear-bound}) at $F=F_{\min}$.
\item[(ii)] Zero-deviation edge: Conversely, there exist symmetric wirings in which the overlaps in Eq.~(\ref{eq:dbar-again}) are canceled cycle by cycle, rendering $f(\phi) = F_{\min}$ for all $\phi$ and hence $D=0$ even at the lower fidelity endpoint. 
\end{itemize}
In both extremes, $F$ can be the same, but the reliability across the inputs---captured solely by $D$---is radically different. This illustrates why $D$ can be an indispensable complement to $F$~\cite{Ghosal2020optimal,Ghosal2021characterizing,Patra2022significance}.

{\em Case 2) Correction asymmetry at fixed $F$.}---Suppose that two protocols, denoted by $P_1$ and $P_2$, are tuned to the same average fidelity $F$ by matching the diagonal trace budget $\sum_\alpha \abs{tr{\bigl(\hat{X}_\alpha\bigr)}}^2$ in Eq.~(\ref{eq:F-closed}). The protocol $P_1$ uses mutually ``scrambled'' corrections (e.g., Heisenberg-Weyl-type sets with approximately uniform pairwise overlaps~\cite{Wootters1989optimal,Gross2007}). On the other hand, the protocol $P_2$ uses ``aligned'' corrections (most $\hat{X}_\alpha$ share an eigen-basis). Because $F$ is insensitive to the pairwise structure, it is identical for both protocols; however, Eq.~(\ref{eq:dbar-again}) shows that the constructive interference of the terms $\abs{\tr{\bigl( \hat{X}_\alpha\hat{X}_\beta \bigr)}}^2$, $\abs{\tr{\bigl( \hat{X}_\alpha\hat{X}_\beta^\dagger \bigr)}}^2$, and
$\tr{\bigl( \hat{X}_\alpha\hat{X}_\beta\hat{X}_\alpha^\dagger\hat{X}_\beta^\dagger \bigr)}$ increases $D$ for the protocol $P_2$, while the scrambled geometry of the protocol $P_1$ minimizes $D$ at fixed $F$. Thus, $(F, D)$ cleanly discriminates the protocols $P_1$ and $P_2$ that would be indistinguishable by $F$ alone.

\section{Teleportation, Bell nonlocality, and the $(F,D)$ witness geometry}\label{sec:nonlocal}

This section turns the abstract link between $d$-dimensional teleportation and Bell nonlocality into concrete via the testable statements on the $(F, D)$ plane. Specifically, we investigate as follow: once the Haar identities of Sec.~\ref{sec:haar} have reduced $F$ and $D$ to trace invariants of the composed unitaries $\{\hat{X}_\alpha\}$, any single snapshot $(F, D)$ can be mapped to a visibility estimate for the isotropic resource, and that visibility can be compared with physically meaningful thresholds---Bell violation and classical simulability.

\subsection{Visibility thresholds in isotropic noise: $p_{\mathrm{BV}}$ and $p_c$, and their $(F,D)$ projections}\label{subsec:cglmp}

We work with the isotropic channel resource $\hat{\rho}_{\mathrm{iso}}(p) = p\ket{\Psi_0}\bra{\Psi_0} + \frac{1-p}{d^2}\hat{\openone}_{d^2}$ introduced in Sec.~\ref{sec:pre}. In a Bell scenario of Collins-Gisin-Linden-Massar-Popescu (CGLMP)~\cite{Collins:2001qdi}, the white-noise contributes no correlations for any fixed choice of local measurements $\mathbb{M}$, while the maximally entangled component contributes fully. Consequently, the CGLMP value scales linearly with the visibility $p$: i.e., 
\begin{eqnarray}
I_d\bigl(\hat{\rho}_{\mathrm{iso}}(p); \mathbb{M}\bigr) = p I_d\bigl(\ket{\Psi_0}; \mathbb{M}\bigr).
\label{eq:Id_linear}
\end{eqnarray}
Optimizing over measurement settings yields
\begin{eqnarray}
I_d^{\mathrm{Q}}(p)=p Q_d, \quad Q_d:=\max_{\mathsf{M}} I_d(\ket{\Psi_0}; \mathbb{M}).
\end{eqnarray}
Since the local hidden-variable (LHV) bound is $I_d^{\mathrm{L}} = 2$, the Bell-violation threshold on the resource visibility is
\begin{eqnarray}
p_{\mathrm{BV}}(d) = \frac{2}{Q_d}.
\label{eq:pBV_def}
\end{eqnarray}
Thus, $p > p_{\mathrm{BV}}(d)$ is necessary and sufficient for CGLMP violation with the isotropic resources. Numerically, it has been found that $p_{\mathrm{BV}}(d=3)\approx 0.696$, $p_{\mathrm{BV}}(d=4)\approx 0.691$, and $p_{\mathrm{BV}}(d)\to 0.673$ as $d\to\infty$~\cite{Collins:2001qdi}. These values will serve as visibility waypoints when we place a nonlocality lines in the $(F, D)$.

A second, the logically independent visibility standard is the separability threshold
\begin{eqnarray}
p_c = \frac{1}{d+1},
\label{eq:pc-again}
\end{eqnarray}
below which $\hat{\rho}_{\mathrm{iso}}(p)$ is separable and above which it is entangled. In our teleportation setting, the classical (measure-and-prepare) benchmark is 
\begin{eqnarray}
F_{\mathrm{cl}}=\frac{2}{d+1}.
\label{eq:F_cl}
\end{eqnarray}
Using the closed form $F_{\max}(p)=p+\frac{1-p}{d}$ from Eq.~(\ref{eq:Fmax}), one checks $F_{\max}(p_c)=F_{\mathrm{cl}}$, so that
\begin{eqnarray}
p > p_c \Longleftrightarrow F_{\max}(p) > \frac{2}{d+1}.
\label{eq:pc-tele-adv}
\end{eqnarray}
In short, $p_c$ marks the onset of the entanglement and the possibility of beating the best classical scheme, whereas $p_{\mathrm{BV}}(d) > p_c$ reveals the strictly stronger regime where the channel resource can also violate a Bell inequality of the CGLMP family. This strict separation of thresholds is the first indication that ``teleportation advantage'' and ``Bell nonlocality'' carve out different aspect~\cite{Popescu:1994,Horodecki1996teleportation,Clifton2001nonlocality} .

To project these visibility conditions to the $(F, D)$ plane, we recall the universal deviation bound in Eq.~(\ref{eq:D-linear-bound}),
\begin{eqnarray}
D \le s_d\bigl(F_{\max}(p)-F\bigr),
\label{eq:D-slope-again}
\end{eqnarray}
where 
\begin{eqnarray}
s_d := \sqrt{\frac{2}{d(d+3)}}.
\end{eqnarray}
These state that, for each fixed $p$, the admissible $(F, D)$ pairs fill a triangle of slope ``$-s_d$'' with the apex at $\bigl(F_{\max}(p), 0\bigr)$ and base at $F_{\min}(p)$. Therefore, we attain:
\begin{itemize}
\item[{\bf (A.1)}] Setting $p = p_{\mathrm{BV}}(d)$ produces a Bell strip in $F$ and, after Eq.~(\ref{eq:D-slope-again}), a ``Bell triangle'' in $(F, D)$. Any point outside this triangle certifies $p > p_{\mathrm{BV}}(d)$, hence a CGLMP-violating resource.
\item[{\bf (A.2)}] Setting $p = p_c$ produces the separability strip in $F$ and its ``classical triangle'' in $(F,D)$, anchored at $\bigl(F_{\max}(p_c),0\bigr) = \bigl(\tfrac{2}{d+1}, 0\bigr)$. The points outside this triangle certify $p > p_c$ and hence the availability of the teleportation advantage.
\end{itemize}

Plotted together, the two triangles (Bell and classical) make the resource landscape visible. The classical region ($p \le p_c$), the entangled-but-local region ($p_c < p \le p_{\mathrm{BV}}$), and the Bell-nonlocal region ($p > p_{\mathrm{BV}}$) appear as nested shapes of increasing $p$, with dimension-dependent slope $s_d$ fixed by Haar-moment identities. These geometric features has briefly been visualized in the qubit (i.e., $d=2$) case~\cite{Bang:2018bbi}. Here, we develop a mathematically rigorous framework for arbitrary $d$ by deriving tight bounds for the pair $(F, D)$ and by making their physical significance in $d$-dimensional teleportation explicit.

\subsection{A Bell-nonlocality witness on the $(F,D)$ plane}\label{subsec:wedge-BV}

We isolate the regime $p \le p_{\mathrm{BV}}(d)$ where no CGLMP violation is possible and characterize all attainable $(F, D)$ in that region. We state the following theorem:
\begin{theorem}[CGLMP-violation wedge]
\label{thm:BV-wedge}
Every isotropic channel resource with $p \le p_{\mathrm{BV}}(d)$ satisfies
\begin{eqnarray}
F &\le& F_{\mathrm{BV}}^{\max}(d) := F_{\max}\bigl(p_{\mathrm{BV}}(d)\bigr) \nonumber \\
D &\le& s_d \bigl(F_{\mathrm{BV}}^{\max}(d) - F\bigr).
\label{eq:FD-at-most-line}
\end{eqnarray}
Thus, all $(F, D)$ from CGLMP-local resources lie in the closed wedge
\begin{widetext}
\begin{eqnarray}
\mathcal{W}_{\mathrm{local}}(d) = \Bigl\{(F,D): F \le F_{\mathrm{BV}}^{\max}(d) ~\mathrm{and}~ D \le s_d\bigl(F_{\mathrm{BV}}^{\max}(d)-F\bigr)\Bigr\}.
\label{eq:FD_wedge}
\end{eqnarray}
\end{widetext}
Hence, any $(F,D)\notin\mathcal{W}_{\mathrm{local}}(d)$ certifies $p>p_{\mathrm{BV}}(d)$ and thus CGLMP nonlocality.
\end{theorem}

The proof is straightforward: From Eq.~(\ref{eq:F-closed}), one has $F \le F_{\max}(p)$ for every $\{\hat{X}_\alpha\}$. Here, if $p \le p_{\mathrm{BV}}(d)$ then $F \le F_{\max}(p)\le F_{\mathrm{BV}}^{\max}(d)$, giving Eq.~(\ref{eq:FD-at-most-line}). Similarly, Eq.~(\ref{eq:D-linear-bound}) gives $D \le s_d\bigl(F_{\max}(p) - F\bigr) \le s_d\bigl(F_{\mathrm{BV}}^{\max}(d) - F\bigr)$, establishing Eq.~(\ref{eq:FD-at-most-line}).

Geometrically, Eq.~(\ref{eq:FD_wedge}) defines a straight exclusion boundary of the slope ``$-s_d$'' that terminates at $(F_{\mathrm{BV}}^{\max}(d), 0)$. Every CGLMP‑local resource (i.e., $p \le p_{\mathrm{BV}}(d)$) must lie on or below this line: at fixed $F$, moving upward in $D$ is forbidden in the local regime. However, once the visibility exceeds the threshold, i.e., $p > p_{\mathrm{BV}}(d)$, the admissible wirings expand beyond this diagonal barrier and the attainable region enters the nonlocal region. The certification proceeds through two complementary (say) doors: either $F > F_{\mathrm{BV}}^{\max}(d)$ or $D > s_d\bigl(F_{\mathrm{BV}}^{\max}(d) - F\bigr)$ with $F \le F_{\mathrm{BV}}^{\max}(d)$. The first door is the ``fidelity‑only'' witness: surpassing the largest average fidelity compatible with $p\le p_{\mathrm{BV}}(d)$ certifies the Bell nonlocality. The second door is genuinely ``deviation‑driven'' witness: even if $F$ sits inside the Bell strip, sufficiently large $D$ still forces the Bell nonlocality. In practice, one reads the universal bound Eq.~(\ref{eq:D-linear-bound}) in reverse: if a measured pair $(F, D)$ lies above the line $D = s_d\bigl(F_{\mathrm{BV}}^{\max}(d) - F\bigr)$, then no choice with $p \le p_{\mathrm{BV}}(d)$ can satisfy the bound, which implies $F_{\max}(p) > F_{\mathrm{BV}}^{\max}(d)$ and hence $p > p_{\mathrm{BV}}(d)$. The linear constraint itself is fixed by the fourth‑moment Haar identity, which reduces $\int d\phi f(\phi)^2$ to the pairwise trace invariants in Eq.~(\ref{eq:dbar-again}). In fact, this is precisely the reason why the wedge in Eq.~(\ref{eq:FD_wedge}) is tight and why $D$ becomes critical witness of the Bell nonlocality.

\subsection{Teleportation advantage: a distinct witness and a distinct threshold}\label{subsec:wedge-adv}

The teleportation shows quantum advantage when it beats the classical limit $F_{\mathrm{cl}}=\frac{2}{d+1}$ [as in Eq.~(\ref{eq:F_cl})]. For isotropic channel resources, the smallest visibility enabling advantage is $p_c=\frac{1}{d+1}$ because $F_{\max}(p_c)=F_{\mathrm{cl}}$ as seen in Eq.~(\ref{eq:pc-tele-adv}).
\begin{theorem}[Teleportation-advantage wedge]
\label{thm:TA-wedge}
Every isotropic channel resource with $p \le p_c$ must satisfy
\begin{eqnarray}
F &\le& F_{\mathrm{sep}}^{\max}(d) := \frac{2}{d+1}, \nonumber \\
D &\le& s_d\bigl(F_{\mathrm{sep}}^{\max}(d) - F\bigr).
\label{eq:FD_classical}
\end{eqnarray}
Equavalently,
\begin{widetext}
\begin{eqnarray}
\mathcal{W}_{\mathrm{cl}}(d) = \Bigl\{(F,D): F \le \tfrac{2}{d+1} ~\mathrm{and}~ D \le s_d\bigl(\tfrac{2}{d+1} - F\bigr)\Bigr\}.
\label{eq:FD_classical_wedge}
\end{eqnarray}
\end{widetext}
Thus, $(F, D) \notin \mathcal{W}_{\mathrm{cl}}(d)$ is necessary for quantum advantage and implies $p>p_c$.
\end{theorem}

The proof is identical to that of {\bf Theorem~\ref{thm:BV-wedge}} with $p_{\mathrm{BV}}(d)$ replaced by $p_c$ and $F_{\mathrm{BV}}^{\max}(d)$ replaced by $F_{\mathrm{sep}}^{\max}(d)$.

In the ideal-correction limit, $\hat{X}_\alpha=\hat{\openone}_d$ for all $\alpha$, so the protocol is perfectly input‑uniform and sits on the horizontal axis $D=0$ with $F=F_{\max}(p)$ [see Eq.~(\ref{eq:Fmax})]. Along this $D=0$ line, the two resource thresholds appear as simple vertical ($F$-line) tests: by Eq.~(\ref{eq:pc-tele-adv}) the onset of the teleportation advantage occurs at $F > F_{\mathrm{cl}}=2/(d+1)$, whereas by Eq.~(\ref{eq:FD-at-most-line}) the onset of the CGLMP-violation requires $F > F_{\mathrm{BV}}^{\max}(d)$. Hence, in this case, the perfect universality ($D=0$) can be quantum yet local whenever $p_c < p \le p_{\mathrm{BV}}(d)$: the protocol beats the classical threshold $F_{\mathrm{cl}}$ without crossing the Bell‑violation boundary.

More generally---away from the $D=0$ axis---the protocol devices can be fluctuated (or tuned) to the same $F$ while exhibiting different $D$. This happens because $F$ depends only on the diagonal trace budget in Eq.~(\ref{eq:F-closed}), namely $\sum_{\alpha}\abs{\tr{\bigl(\hat{X}_\alpha\bigr)}}^{2}$, whereas $D$ is governed by the pairwise invariants collected in Eq.~(\ref{eq:dbar-again}) (e.g., $\abs{\tr{\bigl((\hat{X}_\alpha \hat{X}_\beta\bigr)}}^{2}$, $\abs{\tr{\bigl(\hat{X}_\alpha \hat{X}_\beta^\dagger\bigr)}}^{2}$ and $\tr{\bigl(\hat{X}_\alpha \hat{X}_\beta \hat{X}_\alpha^\dagger \hat{X}_\beta^\dagger\bigr)}$). In this fixed‑$F$ comparison, aligned wirings (many $\hat{X}_\alpha$ sharing an eigen-basis) enlarge those pairwise traces and increase $D$, while scrambled wirings (approximately design‑like sets with weak mutual overlaps) decrease $D$.

With $F$ held equal, a larger measured $D$ forces the point $(F, D)$ to cross the witness lines of {\bf Theorem~\ref{thm:BV-wedge}} and {\bf Theorem~\ref{thm:TA-wedge}} at a smaller horizontal offset, thereby certifying a strictly larger visibility $p$. Formally, this inference is powered by the slope constraint of Eq.~(\ref{eq:D-linear-bound}), which ties $D$ to the gap $F_{\max}(p) - F$: for a fixed $F$, exceeding the line $D = s_d\bigl(F_{\mathrm{BV}}^{\max}(d)-F\bigr)$ (or $D = s_d\bigl(F_{\mathrm{cl}}(d)-F\bigr)$) is possible only if the underlying $F_{\max}(p)$ is larger, and since $F_{\max}(p)$ is monotone in $p$, such a crossing certifies a larger visibility $p$ and thus a stronger resource, even when $F$ alone would not distinguish.

\subsection{Two witnesses, two constraints: $\mathcal{W}_{\mathrm{cl}}(d)$ vs $\mathcal{W}_{\mathrm{local}}(d)$ and why $D$ is decisive}\label{subsec:witness-compare}

To place teleportation advantage and Bell nonlocality on the same footing, we recast both witnesses as linear half‑planes in the $(F,D)$‑plane. Each boundary is a straight line of slope $-s_d$ anchored at a different $D{=}0$ intercept—one at the classical benchmark $F_{\mathrm{cl}}=\tfrac{2}{d+1}$ and the other at the CGLMP cut value $F_{\mathrm{BV}}^{\max}(d)$:
\begin{widetext}
\begin{eqnarray}
\mathcal{W}_{\mathrm{cl}}(d) : F + \frac{D}{s_d} > \frac{2}{d+1}, \qquad \mathcal{W}_{\mathrm{local}}(d) : F + \frac{D}{s_d} > F_{\mathrm{BV}}^{\max}(d)=p_{\mathrm{BV}}(d) + \frac{1-p_{\mathrm{BV}}(d)}{d}.
\label{eq:witness-lines}
\end{eqnarray}
\end{widetext}
Geometrically, these are two parallel ``tilted bars`` slicing the $(F,D)$‑plane. Moving above the lower bar certifies the teleportation advantage; moving above the higher bar certifies the CGLMP-violation. The fact that both bars share the same slope but have different heights is the first sign that $(F,D)$ carries strictly more resolving power than $F$ alone: the vertical gap between the bars is measured directly along the $D$‑axis.

\begin{theorem}[Strict inclusion of witnesses]
\label{thm:inclusion}
For all $d\ge 2$ one has
\begin{eqnarray}
\mathcal{W}_{\mathrm{local}}(d)\subsetneq \mathcal{W}_{\mathrm{cl}}(d).
\label{eq:inclusion}
\end{eqnarray}
\end{theorem}

The proof is simple and as follows. Both lines in Eq.~(\ref{eq:witness-lines}) have the slope $-s_d$. Their $D=0$ intercepts satisfy $F_{\mathrm{BV}}^{\max}(d) > F_{\mathrm{cl}}=\tfrac{2}{d+1}$ because $p_{\mathrm{BV}}(d) > p_c = \tfrac{1}{d+1}$ for all $d \ge 2$. Hence, the nonlocality line lies strictly above the teleportation advantage line, implying Eq.~(\ref{eq:inclusion}).

{\em Fixed‑$F$ vs fixed‑$D$ slices.}---A convenient way to see the distinct constraints is to take the vertical ($F$-fixed) and horizontal ($D$-fixed) cuts.

(i) Fixed $F=F_0$: At any prescribed $F_0$, the two witnesses demand different minimum deviations:
\begin{eqnarray}
D > s_d\bigl(F_{\mathrm{cl}} - F_0\bigr) ~\text{vs.}~ D > s_d\bigl(F_{\mathrm{BV}}^{\max}(d) - F_0\bigr).
\end{eqnarray}
Thus, two protocols fluctuated to the same $F_0$ may fall on opposite sides of the two witnesses because their deviation differ. The protocol with the larger measured $D$ can already certify the nonlocality, while the one with the smaller $D$ certifies at most the quantum advantage. This is a deviation‑enabled separation that $F$ alone cannot provide.

(ii) Fixed $D = D_0 > 0$: Along vertical cuts, we write the average fidelities required to cross each witness as
\begin{eqnarray}
F_{\mathrm{cert}}^{\mathrm{TA}} &=& \frac{2}{d+1} - \frac{D_0}{s_d}, \nonumber \\
F_{\mathrm{cert}}^{\mathrm{BV}} &=& F_{\mathrm{BV}}^{\max}(d) - \frac{D_0}{s_d},
\label{eq:fixedD}
\end{eqnarray}
with $F_{\mathrm{cert}}^{\mathrm{BV}} - F_{\mathrm{cert}}^{\mathrm{TA}} = F_{\mathrm{BV}}^{\max}(d)-\tfrac{2}{d+1} > 0$. Hence, the very same $D_0$ lowers the fidelity requirement by the same amount for both witnesses, but because their anchors differ, the nonlocality threshold remains harder to meet.

{\em Why deviation ($D$), not just mean ($F$), reveals the decisive resource.}---Recall that Eq.~(\ref{eq:F-closed}) shows that $F$ depends only on the diagonal trace budget $\sum_{\alpha}\abs{\tr{\bigl(\hat{X}_\alpha\bigr)}}^2$: it compresses all wiring geometry into a single scalar. By contrast, Eqs.~(\ref{eq:D-cov})--(\ref{eq:dbar-again}) reveal that $D$ is governed by the pairwise invariants $\abs{\tr{\bigl(\hat{X}_\alpha\hat{X}_\beta\bigr)}}^2$, $\abs{\tr{\bigl(\hat{X}_\alpha\hat{X}_\beta^\dagger\bigr)}}^2$, and $\tr{\bigl(\hat{X}_\alpha\hat{X}_\beta\hat{X}_\alpha^\dagger\hat{X}_\beta^\dagger\bigr)}$. This structural split is precisely why $(F, D)$ can discriminate the protocols that $F$ alone cannot: varying the alignment of the composed corrections ${\hat{X}_\alpha}$ can change $D$ without altering $F$.

The universal slope constraint $D \le s_d\bigl(F_{\max}(p) - F\bigr)$ converts any measured deviation into a lower bound on the resource visibility $p$---and hence into a certification tool for both $\mathcal{W}_{\mathrm{cl}}(d)$ and $\mathcal{W}_{\mathrm{local}}(d)$. Put differently: at fixed $F$, only by increasing $D$ one can cross the diagonal barriers in Eq.~(\ref{eq:witness-lines}). This is one of the important messages of our framework: the two wedges impose parallel but distinct linear constraints in $(F, D)$, and the quantity that makes their difference visible is $D$.

\section{Summary and discussions}\label{sec:summary}

We have developed a unified analysis that treats $d$-dimensional teleportation and Bell nonlocality on equal footing by charting both against the paired figures of merit, i.e., the average fidelity $F$ and fidelity deviation $D$. On the technical side, we built a representation-theoretic framework (Schur-Weyl duality plus permutation-symmetry calculus) that reduces all required Haar averages to a finite set of trace invariants of the composed correction unitaries. This yielded the closed-form expressions for $F$ and $D$ for arbitrary Hilbert-space dimension $d$, together with a compact evaluation recipe that is readily reproducible. From these formulas we derived tight and dimension-dependent bounds on fluctuations [see Eq.~(\ref{eq:D-linear-bound}) or Eq.~(\ref{eq:D-slope-again})]. This bound isolates the extent to which non‑uniform performance is forced when the average falls short of the optimal value, and they specialize smoothly to known qubit size.

A single measured point $(F, D)$ could be translated into an estimate of the isotropic-channel visibility, thereby turning the $(F, D)$-plane into a calibrated diagnostic chart with certification lines. In {\bf Theorem~\ref{thm:BV-wedge}} and {\bf Theorem~\ref{thm:TA-wedge}}, we casted the teleportation advantage (relative to $F_{\mathrm{cl}}=2/(d{+}1)$) and the CGLMP nonlocality threshold into two parallel linear witnesses of the identical slope and distinct intercepts; {\bf Theorem~\ref{thm:inclusion}} proved the strict inclusion between the two certified regions. Thus, crossing the lower line guaranteed a teleportation advantage, while crossing the upper line certified the Bell nonlocality. In particular, their vertical separation quantified the physical gap between ``entangled-but-local'' and ``genuinely nonlocal.'' Crucially, $D$ provides an independent route across the Bell boundary: even if $F$ alone lies within the Bell‑local strip, a sufficiently large deviation pushes $(F, D)$ into the nonlocal region, revealing the correlations that a mean-only readout would miss.

Methodologically, the Schur-Weyl-based reduction we develop will furnish a compact calculus for computing high‑dimensional averages and fluctuations well beyond the present setting, e.g., design‑based benchmarking and input‑sensitivity analyses. Our witness-wedge theorems designed by the witness picture on $(F, D)$-plane will provide an operational diagram in which ``entanglement for teleportation'' and ``entanglement for Bell‑violation'' appear as related yet inequivalent. Beyond the present scope, we anticipate several directions where the framework is immediately useful, for example, porting the $(F, D)$ witnesses to one‑sided device‑independent and steering scenarios~\cite{Fan2022quantum}, to alternative Bell families beyond CGLMP~\cite{Son2006generic}, and to non‑isotropic noise models. 

\section*{Acknowledgement}
This work was supported by the Ministry of Science, ICT and Future Planning (MSIP) by the National Research Foundation of Korea (RS-2024-00432214, RS-2025-03532992, and RS-2025-18362970) and the Institute of Information and Communications Technology Planning and Evaluation grant funded by the Korean government (RS-2019-II190003, ``Research and Development of Core Technologies for Programming, Running, Implementing and Validating of Fault-Tolerant Quantum Computing System''), the Korean ARPA-H Project through the Korea Health Industry Development Institute (KHIDI), funded by the Ministry of Health \& Welfare, Republic of Korea (RS-2025-25456722). We acknowledge the Yonsei University Quantum Computing Project Group for providing support and access to the Quantum System One (Eagle Processor), which is operated at Yonsei University.

\onecolumngrid

\appendix

\section{Theoretical Foundations of Haar Averages and Moment Operators}\label{append:A}

In this appendix, we collect several representation-theoretic tools underlying the Haar-integral calculations used throughout the main text. Our goal is to highlight the structural ingredients---rather than to provide full proofs---that justify the formulas appearing in Sec.~\ref{sec:haar} and the subsequent evaluation of the Haar-averaged quantities, such as $F$ and $D$, introduced in Subsec.~\ref{subsec:F} and Subsec.~\ref{subsec:D}. Readers interested in further mathematical details or in broader applications to quantum information may consult Refs.~\cite{Zhang:2014zbq, Ragone:2022axl, Mele2024introductiontohaar}.
The focus here is on the algebraic principles that govern Haar integration, in particular the symmetry constraints imposed by unitary invariance and the resulting decomposition of tensor-power representations.

We begin by recalling the definition and the key invariance properties of the Haar measure on a compact group, in particular on the unitary group $U(d)$, which formalizes the notion of a ``uniformly random'' unitary. These invariance principles imply that the $k$-fold twirling map
\begin{eqnarray}
\mathcal{T}_k(\hat{O}) := \int \hat{U}^{\otimes k} \hat{O} (\hat{U}^\dagger)^{\otimes k} dU
\end{eqnarray}
is equivariant with respect to the tensor-power action of $U(d)$, and hence its image must lie in the commutant of this action. In other words, Haar integration acts as the projection onto the algebra of operators that commute with $\hat{U}^{\otimes k}$. By the Schur-Weyl duality, this commutant is spanned by permutation operators, which explains their general appearance in Haar-averaged expressions and why the quantities, such as twirls, random fidelities, and pure-state moments reduce to simple invariant tensors.

Building on this foundation, we introduce the $k$-th moment operators, which encode the action of the Haar measure on the $k$-fold tensor-power (fundamental) representation $\hat{U}^{\otimes k}$ acting on $(\mathbb{C}^d)^{\otimes k}$. These operators are determined by the commutant of $\hat{U}^{\otimes k}$ and hence admit an explicit expansion in the permutation basis given by the Schur-Weyl duality. To obtain the coefficients in this expansion, we employ the Weingarten calculus, which provides closed-form formulas for integrals
\begin{eqnarray}
\int \hat{U}_{i_1 j_1} \cdots \hat{U}_{i_k j_k} \overline{\hat{U}}_{i'_1 j'_1} \cdots \overline{\hat{U}}_{i'_k j'_k} dU = \sum_{\sigma,\tau \in S_k}  \delta_{i_1, i'_{\sigma(1)}} \cdots \delta_{i_k, i'_{\sigma(k)}} \delta_{j_1, j'_{\tau(1)}} \cdots \delta_{j_k, j'_{\tau(k)}} Wg(\sigma^{-1}\tau , d),
\end{eqnarray}
and, more generally, for arbitrary $k$-fold twirls and Haar moments of pure states. In this sense, the Weingarten calculus serves as the computational companion to the structural decomposition: the Schur-Weyl duality determines the invariant tensor basis, while the Weingarten functions supply the coefficients that make the expressions fully explicit. For convenience, we also compile several trace identities and cycle-factorization rules that streamline the evaluation of the Haar-averaged quantities appearing in our main text.

\subsection{Haar measure}

The Haar measure on a compact group provides the unique notion of uniformity compatible with the group structure. For the unitary group $U(d)$, we denote this measure by $d\mu_H(U)$ or simply $dU$.
\begin{definition}
Let $G$ be a compact group. A probability measure $\mu_H$ on $G$ is called the \emph{Haar measure} if, for every measurable set $S \subseteq G$ and integrable function $f:G \to \mathbb{C}$,
\begin{eqnarray}
\mu_H(S)\ge 0, \quad \mu_H(G)=1,
\end{eqnarray}
and
\begin{eqnarray}
\mathbb{E}_{U\sim\mu_H}[f(U)] := \int_G f(U) d\mu_H(U).
\end{eqnarray}
\end{definition}
The defining feature of the Haar measure is its invariance under the group action.
\begin{proposition}[Invariance properties]
For all $U, V \in U(d)$ and every integrable $f$,
\begin{eqnarray}
\int f(VU) dU &=& \int f(U) dU, \nonumber \\
\int f(UV) dU &=& \int f(U) dU, \nonumber \\
\int f(VUV^\dagger) dU &=& \int f(U) dU, \nonumber \\
\int f(U^\dagger) dU &=& \int f(U) dU.
\end{eqnarray}
\end{proposition}

These identities express that the Haar measure on $U(d)$ is left- and right-invariant and hence unique. The crucial consequence of this invariance is the following: the $k$-fold twirling map $\mathcal{T}_k(\hat{O}) := \int \hat{U}^{\otimes k} \hat{O} (\hat{U}^\dagger)^{\otimes k} dU$ is \emph{equivariant} with respect to the tensor-power representation of $U(d)$. Therefore, $\mathcal{T}_k(\hat{O})$ must lie in the commutant of $\hat{U}^{\otimes k}$ and admits a decomposition in the permutation-operator basis by the Schur-Weyl duality.

\subsection{$k$-th moment operators}

To describe the averaged action of Haar-random unitaries on $k$-copy systems, we introduce the $k$-th moment operator, which formalizes the $k$-fold unitary twirl over the unitary group.
\begin{definition}[$k$-th moment operator / $k$-fold unitary twirl]
For any $k \in \mathbb{N}$, the $k$-th moment operator
\begin{eqnarray}
\mathcal{T}_k : \mathcal{L}((\mathbb{C}^d)^{\otimes k}) \longrightarrow \mathcal{L}((\mathbb{C}^d)^{\otimes k})
\end{eqnarray}
is defined by
\begin{eqnarray}
\mathcal{T}_k({\mathcal O}) := \mathbb{E}_{{U}\sim\mu_H} \bigl[ \hat{U}^{\otimes k} \hat{O} \hat{U}^{\dagger \otimes k} \bigr], ~~\hat{O}\in\mathcal{L}((\mathbb{C}^d)^{\otimes k}).
\end{eqnarray}
\end{definition}
The map $\mathcal{T}_k$ characterizes the averaged action of conjugating an operator by $k$-fold tensor powers of Haar-random unitaries, and thereby describes how random unitaries act on multi-copy systems. By Haar invariance, $\mathcal{T}_k$ is linear, trace-preserving, and self-adjoint with respect to the Hilbert-Schmidt inner product. Moreover, $\mathcal{T}_k$ is idempotent and hence acts as an orthogonal projection onto the subspace of operators that are invariant under simultaneous conjugation by $\hat{U}^{\otimes k}$ for all $\hat{U} \in U(d)$. This projection structure is precisely what allows Haar averages to reduce to symmetry-invariant components.

The $k$-th moment operator also provides a natural language for quantifying how well a finite ensemble of unitaries mimics the statistical properties of the Haar measure. Indeed, many notions of pseudo-randomness in quantum information are defined by requiring that an ensemble reproduce Haar moments up to some order. This leads to the concept of a \emph{unitary $t$-design}~\cite{Dankert:2009,Gross2007}, which is an ensemble that matches the Haar moment operators through order $t$.
\begin{definition}[Unitary $t$-design via moment operators]
A finite ensemble of unitaries $\mathcal{E}=\{{U}_j\}$ is an \emph{exact unitary $t$-design} if, for all $k \le t$ and all $\hat{O} \in \mathcal{L}((\mathbb{C}^d)^{\otimes k})$,
\begin{eqnarray}
\frac{1}{\abs{\mathcal{E}}} \sum_{\hat{U}_j\in\mathcal{E}}\hat{U}_j^{\otimes k}\hat{O}\hat{U}_j^{\dagger\otimes k} = \mathcal{T}_k(\hat{O}).
\end{eqnarray}
\end{definition}
Thus, the moment operator serves as the benchmark against which finite ensembles are compared: an ensemble forms a $t$-design precisely when it reproduces the Haar moments up to order $t$.

\subsection{Schur-Weyl duality and the structure of Haar averages}

The analysis of the $k$-th moment operator reduces to understanding the operators that remain invariant under conjugation by $\hat{U}^{\otimes k}$ for all $\hat{U} \in U(d)$. This invariant subalgebra is precisely the \emph{commutant} of the $k$-fold tensor-power representation, and it governs the entire structure of Haar averages. Before invoking representation-theoretic tools, we formalize this object.
\begin{definition}[$k$-th order commutant]
For a set of operators $S \subset \mathcal{L}(\mathbb{C}^d)$, the $k$-th order commutant is
\begin{eqnarray}
\mathrm{Comm}(S, k) := \left\{ \hat{A} \in \mathcal{L}((\mathbb{C}^d)^{\otimes k}) : \bigl[\hat{A}, \hat{B}^{\otimes k} \bigr] = 0 ~ \bigl(\forall \hat{B} \in S \bigr) \right\}.
\end{eqnarray}
This is a subalgebra of $\mathcal{L}((\mathbb{C}^d)^{\otimes k})$, and it consists of the operators that are fixed by the Haar twirl.
\end{definition}

Because the Haar twirl $\mathcal{T}_k$ is an orthogonal projection onto this commutant, any operator $\hat{O}$ admits the expansion
\begin{eqnarray}
\mathcal{T}_k(\hat{O}) = \sum_{i=1}^{\dim G_{k}} \langle \hat{P}_i , \hat{O} \rangle_{HS} \hat{P}_i,
\label{eq:Tk_comm_expansion}
\end{eqnarray}
where $G_k = \mathrm{Comm}(U(d), k)$ and $\{ \hat{P}_i \}$ is an orthonormal basis with respect to the Hilbert-Schmidt inner product. Thus the problem of determining Haar averages is reduced to identifying an explicit basis for $G_k$.

A fundamental structural fact is that, when $S = U(d)$, the commutant algebra takes an extremely simple and concrete form. The key structural result is the Schur-Weyl duality, which decomposes the $k$-fold tensor power of the defining representation and simultaneously determines its commutant. We state it as the following theorem.
\begin{theorem}[Schur-Weyl duality]
For the natural representation of $U(d)$ on $(\mathbb{C}^d)^{\otimes k}$,
\begin{eqnarray}
\mathrm{Comm}(U(d), k) = \mathrm{span}\bigl\{ \hat{V}_d(\pi) : \pi \in S_k \bigr\},
\end{eqnarray}
where $\hat{V}_d(\pi)$ denotes the permutation operator corresponding to $\pi \in S_k$ acting on tensor factors.
\end{theorem}
The Schur-Weyl duality tells us that the permutation operators completely characterize all Haar-invariant tensors on $k$ copies. As a consequence, the expansion in Eq.~(\ref{eq:Tk_comm_expansion}) becomes explicit:
\begin{eqnarray}
\mathcal{T}_k(\hat{O}) = \sum_{\pi\in S_k} t_\pi \hat{V}_d(\pi), \quad t_\pi \in \mathbb{C}.
\label{eq:Tk_perm_expansion}
\end{eqnarray}
In this form, the structural constraints imposed by Haar invariance become transparent: the Haar measure \emph{forgets} everything except the components of $\hat{O}$ that lie along permutation symmetries of the tensor-product space.

What remains is the explicit evaluation of the coefficients $t_\pi$. These coefficients are determined by the Weingarten calculus, which inverts the Gram matrix of permutation operators and yields closed-form expressions for integrals of $\hat{U}$ and $\hat{U}^\dagger$. In this way, the Schur-Weyl duality provides the invariant tensor basis, while the Weingarten calculus provides the weights that complete the Haar average.

\subsection{Permutation operators and invariant projectors}\label{sec:perm_proj}

The permutation representation of the symmetric group plays a central role in rendering the expansion in Eq.~(\ref{eq:Tk_perm_expansion}) concrete. Since the Schur-Weyl duality identifies the permutation operators as a basis for the Haar-invariant commutant, we collect here the algebraic properties that make these operators effective tools for decomposing tensor-power Hilbert spaces.

\begin{definition}[Permutation operators]
For $\pi \in S_k$, the corresponding permutation operator on $(\mathbb{C}^d)^{\otimes k}$ is
\begin{eqnarray}
\hat{V}_d(\pi) = \sum_{j_1,\ldots,j_k=0}^{d-1} \ket{j_{\pi^{-1}(1)}, j_{\pi^{-1}(2)}, \ldots, j_{\pi^{-1}(k)}}\bra{j_1,j_2,\ldots,j_k}.
\label{eq:def_perm_operator}
\end{eqnarray}
Here, the inverse appears because we adopt the convention that operators act on kets from the left.
\end{definition}
This representation acts by permuting the tensor factors, and it extends linearly to a unitary representation of $S_k$. The permutation operators satisfy the following properties:
\begin{eqnarray}
&& \hat{V}_d(\mathrm{id}) = \hat{\openone}, \nonumber \\
&& \hat{V}_d(\pi) \hat{V}_d(\nu) = \hat{V}_d(\pi\nu), \nonumber \\
&& \hat{V}_d(\pi)^\dagger = \hat{V}_d(\pi^{-1}), \nonumber \\
&& \hat{V}_d(\pi)\bigl(\ket{\psi_1}\otimes\cdots\otimes \ket{\psi_k}\bigr) = \ket{\psi_{\pi^{-1}(1)}}\otimes\cdots\otimes \ket{\psi_{\pi^{-1}(k)}}, \nonumber \\
&& \hat{V}_d(\pi) \bigl( \hat{A}_1 \otimes\cdots\otimes \hat{A}_k \bigr) \hat{V}_d(\pi)^\dagger = \hat{A}_{\pi^{-1}(1)} \otimes\cdots\otimes \hat{A}_{\pi^{-1}(k)}.
\label{eq:perm_properties}
\end{eqnarray}
Thus, for an operator $\hat{A}_{1\cdots k}$ acting on $k$ subsystems, conjugation by $\hat{V}_d(\pi)$ effects a reindexing:
\begin{eqnarray}
\hat{V}_d(\pi) \hat{A}_{1\cdots k}  \hat{V}_d(\pi)^{-1} = \hat{A}_{\pi^{-1}(1) \ldots \pi^{-1}(k)}.
\label{eq:perm_conjugation}
\end{eqnarray}

A basic characteristic feature of the permutation operators is that their traces depend only on the cycle structure of the permutation.
\begin{proposition}[Trace formula]
\label{traceformula}
For $\sigma,\pi \in S_k$, 
\begin{eqnarray}
\tr{\bigl(  \hat{V}_d(\sigma) \hat{V}_d(\pi) \bigr)} = \tr{\bigl( \hat{V}_d(\sigma\pi) \bigr)} = d^{\#\mathrm{cyc}(\sigma\pi)},
\label{eq:trace_formula_perm}
\end{eqnarray}
where the trace depends only on the cycle structure of the composed permutation, and $\#\mathrm{cyc}(\rho)$ denotes the number of disjoint cycles in $\rho \in S_k$.
\end{proposition}
In particular, the trace is maximal ($d^k$) when $\pi=\sigma^{-1}$ and takes the value $d^c$ where $c \leq k$ is the number of cycles. This cycle-counting rule is an essential ingredient in computing explicit forms of $k$-th moments and appears repeatedly in Haar-average calculations.

Beyond their algebraic properties, permutation operators enable the construction of invariant projectors onto symmetry sectors of the tensor-product space. The simplest examples are the symmetric and antisymmetric subspaces, which correspond to the trivial and sign irreducible representations of $S_k$.
\begin{definition}[Symmetric subspace]
The symmetric subspace is
\begin{eqnarray}
\mathrm{Sym}_k(\mathbb{C}^d) = \left\{ \ket{\psi} : \hat{V}_d(\pi)\ket{\psi}=\ket{\psi} \bigl(\forall \pi \in S_k \bigr) \right\},
\end{eqnarray}
and it is the image of the projector
\begin{eqnarray}
\hat{P}_{\mathrm{sym}}^{(d,k)} = \frac{1}{k!} \sum_{\pi\in S_k} \hat{V}_d(\pi),
\label{eq:sym_proj}
\end{eqnarray}
which satisfies $\hat{P}_{\mathrm{sym}}^{(d,k)}{}^{2} = \hat{P}_{\mathrm{sym}}^{(d,k)}$ and $\hat{P}_{\mathrm{sym}}^{(d,k)}{}^{\dagger} = \hat{P}_{\mathrm{sym}}^{(d,k)}$.
\end{definition}

\begin{definition}[Antisymmetric subspace]
The antisymmetric subspace is
\begin{eqnarray}
\mathrm{ASym}_k(\mathbb{C}^d) = \left\{ \ket{\psi}: \hat{V}_d(\pi)\ket{\psi} = \mathrm{sgn}(\pi)\ket{\psi}, ~~ \bigl( \forall \pi \in S_k \bigr) \right\},
\end{eqnarray}
and it is the image of the projector
\begin{eqnarray}
\hat{P}_{\mathrm{asym}}^{(d,k)} = \frac{1}{k!} \sum_{\pi\in S_k} \mathrm{sgn}(\pi)\, {V}_d(\pi).
\label{eq:asym_proj}
\end{eqnarray}
which satisfies $ {P}_{\mathrm{asym}}^{(d,k)}{}^{2}= {P}_{\mathrm{asym}}^{(d,k)}$ and
$ {P}_{\mathrm{asym}}^{(d,k)}{}^{\dagger}= {P}_{\mathrm{asym}}^{(d,k)}$.
\end{definition} 
Note that the symmetric and antisymmetric projectors are orthogonal,
\begin{eqnarray}
\hat{P}_{\mathrm{asym}}^{(d,k)} \hat{P}_{\mathrm{sym}}^{(d,k)} = 0,
\end{eqnarray}
and together they illustrate how permutation operators resolve the tensor-power space into invariant components. More generally, the Schur-Weyl duality states that all irreducible subspaces arise from analogous projectors built from the characters of $S_k$. Thus, the permutation operators introduced above form the basic algebraic building blocks for the decomposition used in evaluating Haar moments.

\subsection{$k$-th moment and Weingarten expansion}

The Schur-Weyl duality implies that all operators commuting with the collective action ${U}^{\otimes k}$ of the unitary group $U(d)$ form the algebra spanned by the permutation operators $\{ \hat{V}_d(\pi)\}_{\pi\in S_k}$. This structural fact is the basis of the Weingarten calculus, which expresses Haar averages as weighted sums over permutations, with weights given by the Weingarten function~\cite{Collins:2006jgn}.

\emph{Expansion of the moment operator.}---For any operator $\hat{O}$ acting on $(\mathbb{C}^d)^{\otimes k}$, its $k$-th moment under the Haar measure can be written in the permutation basis as
\begin{eqnarray}
\mathcal{T}_k(\hat{O}) = \sum_{\pi\in S_k} t_\pi \hat{V}_d(\pi), \quad t_\pi \in \mathbb{C}.
\label{eq:moment_perm_expansion_new}
\end{eqnarray}
To determine the coefficients $\{t_\pi\}$, we choose a fixed $\sigma \in S_k$, multiply Eq.~(\ref{eq:moment_perm_expansion_new}) by $\hat{V}_d(\sigma^{-1})$ on the right, and take the trace. Using the orthogonality relation from Proposition~\ref{traceformula}, we obtain
\begin{eqnarray}
\tr{\bigl[\mathcal{T}_k(\hat{O})\hat{V}_d(\sigma^{-1})\bigr]} = \sum_{\pi \in S_k} t_\pi d^{\#\mathrm{cyc}(\pi\sigma^{-1})}.
\label{eq:linear_system_new}
\end{eqnarray}
Since $\hat{V}_d(\sigma^{-1})$ commutes with $\hat{U}^{\otimes k}$, the left-hand side depends only on $\hat{O}$:
\begin{eqnarray}
y_\sigma :=  \tr{\bigl[ \hat{O} \hat{V}_d(\sigma^{-1}) \bigr]}.
\label{eq:def_y_sigma_new}
\end{eqnarray}
Introducing the Gram matrix $G^{(k)}_{\pi,\sigma} := d^{\#\mathrm{cyc}(\pi\sigma^{-1})}$, Eq.~(\ref{eq:linear_system_new}) gives the linear system 
\begin{eqnarray}
G^{(k)} t = y.
\end{eqnarray}
Inverting $G^{(k)}$ yields the Weingarten expansion:
\begin{eqnarray}
\mathcal{T}_k( \hat{O}) = \sum_{\pi,\sigma \in S_k} \mathrm{Wg}(\pi\sigma^{-1}, d) \tr{\bigl[ \hat{O} \hat{V}_d(\sigma^{-1})\bigr]} \hat{V}_d(\pi),
\label{eq:weingarten_new}
\end{eqnarray}
where $\mathrm{Wg}(\cdot,d)$ denotes the unitary Weingarten function.

\emph{Generalization to compact groups.}---The above derivation extends to any compact group $G$. Let $G_k = \mathrm{Comm}(G,k)$ denote the commutant of the $k$-fold representation of $G$, and choose a basis $\{ \hat{P}_\pi\}_{\pi \in G_k}$ of this algebra. Then, the $k$-th moment takes the form:
\begin{theorem}
Let $G$ be compact.  
For $ {\mathcal O}$ acting on the $k$-fold representation space,
\begin{eqnarray}
\mathcal{T}_k( \hat{O}) = \sum_{\pi,\sigma \in G_k} \mathrm{Wg}(\pi\sigma^{-1}, d) \tr{\bigl[ {\hat{O} \hat{P}_{\sigma^{-1}} }\bigr]} \hat{P}_\pi,
\end{eqnarray}
where $\mathrm{Wg}(\cdot,d)$ is the Weingarten function associated with the compact group $G$.
\label{thm:weingarten_general_new}
\end{theorem}

\subsection{Weingarten calculus and trace identities}

Building on the general Weingarten expansion for compact groups established in Theorem~\ref{thm:weingarten_general_new}, we now focus on the unitary group $U(d)$. We extract several fundamental identities for its Weingarten function and the relevant trace decompositions, which provide the main computational tools for the Haar averages used in this work.

Before presenting the identities themselves, let us recall that the Weingarten function is defined as the inverse of the Gram matrix $G^{(k)}_{\pi,\sigma}=d^{\#\mathrm{cyc}(\pi\sigma^{-1})}$. As a result, the Weingarten function inherits several nontrivial summation and convolution properties reflecting the cycle structure of the symmetric group. The following lemmas capture two such properties that will repeatedly appear later.

\emph{Basic identities for the unitary Weingarten function.}---The first lemma provides a closed-form expression for the total sum of Weingarten coefficients. Although the identity may seem somewhat opaque at first sight, it follows directly from the way in which the Gram matrix encodes cycle-counting information.
\begin{lemma}
For the Weingarten function of $U(d)$,
\begin{eqnarray}
\sum_{\tau \in S_k} \mathrm{Wg}(\tau, d) = \frac{1}{\prod_{i=0}^{k-1} (d+i)}.
\end{eqnarray}
\label{lem:wg_sum_new}
\end{lemma}
This identity is an immediate consequence of the fact that the Gram matrix satisfies
\begin{eqnarray}
\sum_{\sigma\in S_k} d^{\#\sigma} = \prod_{i=0}^{k-1}(d+i),
\end{eqnarray}
a classical generating-function identity for the Stirling numbers of the first kind. Intuitively, the lemma reflects that a uniform summation of the inverse Gram matrix returns the reciprocal of the total cycle-weight contribution encoded in the Gram matrix.

A second important identity concerns the convolution of the Weingarten function with an arbitrary class function on $S_k$. This lemma shows that such a convolution completely factorizes, which drastically simplifies many later expressions involving moment operators.
\begin{lemma}
For any function $f:S_k\to\mathbb{C}$,
\begin{eqnarray}
\sum_{\pi,\sigma\in S_k} \mathrm{Wg}(\pi^{-1}\sigma, d) f(\sigma) = \sum_{\pi\in S_k} \mathrm{Wg}(\pi, d) \times \sum_{\tau\in S_k} f(\tau).
\end{eqnarray}
\label{lem:wg_convolution_new}
\end{lemma}
This result is essentially a statement that the Weingarten function is a central element in the convolution algebra of functions on $S_k$. Because of this factorization, whenever a Haar average produces a sum over permutations weighted by traces, one can immediately decouple the Weingarten part from the trace part.

Combining Lemma~\ref{lem:wg_sum_new} and Lemma~\ref{lem:wg_convolution_new}, we obtain a compact expression for the \emph{entry-wise} Haar average of the moment operator in the computational basis.
\begin{proposition}
\label{prop:k_moment_contraction_new}
For any operator $\hat{O}$,
\begin{eqnarray}
\bra{0}^{\otimes k} \mathcal{T}_k( \hat{O}) \ket{0}^{\otimes k} = \frac{1}{d(d+1)\cdots(d+k-1)} \sum_{\pi\in S_k} \tr{\bigl[ {V}_d^\dagger(\pi) \hat{O} \bigr]}.
\end{eqnarray}
\end{proposition}
This proposition is particularly useful in the main text, where the evaluation of entanglement-related quantities reduces to computing such matrix elements of the Haar-twirled operator.

\emph{Haar averages of tensor powers of pure states.}---As a first application, recall that the Haar twirling of the operator $\bigl( \ket{\phi}\bra{\phi} \bigr)^{\otimes k}$ under the $k$-fold tensor representation of $U(d)$ produces the normalized projector onto the symmetric subspace. The Weingarten calculus gives a direct and elementary proof:
\begin{proposition}
\label{prop:pure_state_moment_new}
For any $\ket{\phi} \in \mathbb{C}^d$ and any integer $k \ge 1$,
\begin{eqnarray}
\mathbb{E}_{ {U}\sim\mu_H} \left[ \hat{U}^{\otimes k} (\ket{\phi}\bra{\phi})^{\otimes k} \hat{U}^{\dagger\otimes k} \right] = \frac{\hat{P}_{\mathrm{sym}}^{(d,k)}}{\tr{ \bigl[ \hat{P}_{\mathrm{sym}}^{(d,k)} \bigr]}},
\end{eqnarray}
where $
\tr{\bigl[ \hat{P}_{\mathrm{sym}}^{(d,k)} \bigr]} = {{k+d-1} \choose {k}}$.
\end{proposition}
Here, the permutation invariance of $(\ket{\phi}\bra{\phi})^{\otimes k}$ ensures that the Weingarten expansion contributes solely to the symmetric subspace, resulting in the symmetric projector.

\emph{Trace identities contracted with permutation operators.}---The remaining ingredient needed for explicit Haar-average calculations is the ability to evaluate traces of operator tensors contracted with permutation operators, i.e.,
\begin{eqnarray}
\tr{\left[ \bigl( \hat{A}_1 \otimes\cdots\otimes \hat{A}_k \bigr) \hat{V}_d(\pi) \right]}. 
\end{eqnarray}
This identity is the backbone of explicit $k$-th moment evaluations. It reveals a direct correspondence between the cycle structure of $\pi$ and products of the operators $\{ \hat{A}_j \}$.
\begin{proposition}
Let $\pi \in S_k$ with disjoint cycle decomposition
\begin{eqnarray}
\pi = c_1 c_2 \cdots c_r,
\end{eqnarray}
where each cycle $c=(l_1\,l_2\,\dots\,l_{k_c})$ has length $k_c$. Then, for operators $\hat{A}_1 ,\ldots, \hat{A}_k \in \mathcal{L}(\mathbb{C}^d)$, the trace has the following decomposition:
\begin{eqnarray}
\tr{\left[ \bigl( \hat{A}_1 \otimes\cdots\otimes \hat{A}_k \bigr) \hat{V}_d(\pi) \right]} = \prod_{c \in \pi} \tr{\left[ \prod_{m=0}^{k_c-1} \hat{A}_{c^{-m} (l_c)} \right]}.
\end{eqnarray}
\end{proposition}
Thus, the trace factorizes over the cycles of $\pi$, with each cycle contributing a product of operators in cyclic order. Together with the Weingarten identities, this trace factorization completes the set of tools required for all Haar averages appearing in the main text.

\twocolumngrid

\bibliography{refs_QTdev}

\end{document}